\documentclass[osajnl,twocolumn,showpacs,superscriptaddress,10pt]{revtex4-1}

\usepackage{mathtools}
\usepackage{amsmath,amssymb,graphicx}
\usepackage[lofdepth,lotdepth,caption=false]{subfig}
\usepackage{textcomp}

\begin{document}

\title{Orthogonality breaking sensing model based on the instantaneous Stokes vector and the Mueller calculus}

\author{No\'e Ortega-Quijano}\email{Corresponding author: noe.ortega@univ-rennes1.fr}
\affiliation{Institut de Physique de Rennes, CNRS, Universit\'e de Rennes 1, Campus de Beaulieu, 35\,042 Rennes, France}

\author{Julien Fade}
\affiliation{Institut de Physique de Rennes, CNRS, Universit\'e de Rennes 1, Campus de Beaulieu, 35\,042 Rennes, France}

\author{Muriel Roche} 
\affiliation{\mbox{Aix-Marseille Universit\'e, CNRS, Centrale Marseille, Institut Fresnel, UMR 7249, 13\,013 Marseille, France}}

\author{Fran\c{c}ois Parnet}
\affiliation{Institut de Physique de Rennes, CNRS, Universit\'e de Rennes 1, Campus de Beaulieu, 35\,042 Rennes, France}

\author{Mehdi Alouini}
\affiliation{Institut de Physique de Rennes, CNRS, Universit\'e de Rennes 1, Campus de Beaulieu, 35\,042 Rennes, France}

\begin{abstract}

Polarimetric sensing by orthogonality breaking has been recently proposed as an alternative technique for performing direct and fast polarimetric measurements using a specific dual-frequency dual-polarization (DFDP) source. Based on the instantaneous Stokes-Mueller formalism to describe the high-frequency evolution of the DFDP beam intensity, we thoroughly analyze the interaction of such a beam with birefringent, dichroic and depolarizing samples. This allows us to confirm that orthogonality breaking is produced by the sample diattenuation, whereas this technique is immune to both birefringence and diagonal depolarization. We further analyze the robustness of this technique when polarimetric sensing is performed through a birefringent waveguide, and the optimal DFDP source configuration for fiber-based endoscopic measurements is subsequently identified. Finally, we consider a stochastic depolarization model based on an ensemble of random linear diattenuators, which makes it possible to understand the progressive vanishing of the detected orthogonality breaking signal as the spatial heterogeneity of the sample increases, thus confirming the insensitivity of this method to diagonal depolarization. The fact that the orthogonality breaking signal is exclusively due to the sample dichroism is an advantageous feature for the precise decoupled characterization of such an anisotropic parameter in samples showing several simultaneous effects.
	
\end{abstract}


\maketitle

\section{Introduction}
Polarimetric imaging systems are of growing interest for many
applications like remote sensing \cite{tyo2006}, astronomy
\cite{tingerben1996}, defense \cite{goudail2002,demos1997},
biomedicine \cite{smith00,bueno2007,ghosh11,pierangelo11}, synthetic-aperture radar
\cite{lee2001}, and machine or enhanced vision
\cite{meriaudeau2008,fade2014}. Active polarimetric techniques allow
valuable information of a given scene or sample to be retrieved from
the measure of its anisotropic and depolarizing properties. Mueller
polarimetry is the most exhaustive active technique, as it completely
characterizes the polarimetric parameters of an object. This technique
typically implies registering 16 images using different approaches to
modulate the measurement in the spatial, temporal, and/or spectral
domain \cite{tyo2002,demartino2003,jiao2005,lacasse2011,alenin2014,legratiet2015}. However, the
complexity of the system and the loss of performance entailed by the
different multiplexing methods have led to propose several simplified
polarimetric techniques which optimize the measurement of some
specific polarimetric properties at a high performance
\cite{anna2012,tower2001,refregier2007,nan2009}.

In this context, the implementation of fiber-guided polarimetric
imaging systems constitutes a challenging issue, due to the fact that
the state of polarization of the illuminating beam is modified by the
optical waveguide in an uncontrolled way. This is a remarkable aspect
to be adressed for endoscopic applications, where the optical fiber
stress-induced birefringence is the dominant effect on beam polarization
\cite{wood2010}. The feasibility of a multimodal endoscopic system
including cross-polarized imaging has been demonstrated for Barett's
metaplasia imaging \cite{thekkek2013}. However, such a technique
provides an orientation-dependent contrast, which entails some
drawbacks for \emph{in vivo} applications. Moreover, the fact that the
polarimetric elements and the CCD camera are placed in the distal end
of the endoscope is quite restrictive in terms of
miniaturization. Regarding Mueller polarimetry, a narrow band
3$\times$3 Mueller polarimetric endoscope was presented and validated
\emph{ex vivo} on a Sprague-Dawley rat \cite{qi2013}. However, the use
of a rigid endoscope is unfeasible for most practical
applications. Finally, a full Mueller endoscopic polarimeter, based on a first characterization of the optical fiber using a micro-switchable mirror before every Mueller matrix acquisition, was proposed in \cite{manhas2015}. Recently, this novel technique has been combined with a spectral encoding of polarimetric channels to significantly reduce the acquisition time \cite{vizet2015}.

Recently, we proposed a novel polarimetric imaging modality based on
the orthogonality breaking sensing principle \cite{fade12}. This
technique uses a dual-frequency dual-polarization (DFDP) coherent
source, and is based on the measurement of the detected intensity
component at the radio-frequency beatnote after interaction with the
sample or scene under analysis. This technique enables a subset of its
polarimetric properties to be determined from a single acquisition at
both high speed and high dynamic range. The first implementation of a
polarimetric contrast microscope by orthogonality breaking was
presented in \cite{schaub14}. A modified source architecture, enabling
the linear diattenuation and optic axis of the sample to be completely characterized, 
was subsequently described and validated in
\cite{ortega15}. Though insensitive to birefringence effects which are predominant in biological tissues, this technique provides an alternative method to characterize biological anisotropic structures. Indeed, biological tissues usually show both birefringence and dichroism sharing the same anisotropy axis \cite{jiao2003,park2004}. As a result, biological anisotropic structures can nonetheless be addressed by measuring their diattenuation properties, provided the measurement dynamics is high enough. However, all these previous works on orthogonality breaking imaging \cite{fade12,schaub14,ortega15} were focused on specific examples and particular measurement configurations. The in-depth analysis
of the physical origin of the orthogonality breaking signal amplitude
and phase, in relation with the polarimetric properties of the sample
and the characteristics of the illumination system, is still an open question.

In this work, we develop a comprehensive theoretical model of
orthogonality breaking sensing based on the instantaneous Stokes
vector and the Mueller calculus. This approach makes it possible to
describe the interaction of the DFDP beam with anisotropic
depolarizing samples. Based on this method, we develop a thorough
analysis of the orthogonality breaking signal characteristics for both
free-space and fiber-guided measurements. The results are presented
for the basic types of optical elements, namely isotropic absorbers,
elliptical, circular, and linear retarders and diattenuators, and
diagonal depolarizers. This theoretical analysis is then completed
with a detailed discussion, based on both simulations and experimental
measurements, which allows us to conclude that the
orthogonality-breaking technique is definitely not sensitive to diagonal
depolarization. Throughout this work, a special emphasis is made on
the practical implications of each configuration for experimental
polarimetry.

This paper is organized as follows: the instantaneous Stokes vector
description of a polarized light beam is reviewed in
Section~\ref{sec:stokes}, before applying it to the specific
characterization of the DFDP source in Section~\ref{sec:dfdp}. Using the instantaneous
Stokes-Mueller formalism, we then thoroughly study in
Section~\ref{sec:freespaceob} how this type of laser source can be
used for free-space polarimetric sensing by the orthogonality breaking
principle. The influence of an optical waveguide on orthogonality
breaking sensing is then investigated in
Section~\ref{sec:endoscopicob} to analyze the potential of this
technique for endoscopic applications. Lastly, a discussion on the
sensitivity of this technique to depolarization is presented in
Section~\ref{sec:discussion} using a stochastic model of linear
diattenuation with random optic axis orientation, before the
conclusion of this work is given in Section~\ref{sec:concl}.

\section{Instantaneous Stokes vector}\label{sec:stokes}
Firstly, we consider a fluctuating optical plane wave described by its
transverse electric field. The reference frame is set so that
propagation is along the $z$ axis in a right-handed Cartesian
coordinate system $xyz$, and thus the complex electric field at a
specific point $z_{0}$ can be resolved into a pair of orthogonal
polarization states:
\begin{equation}
\vec{\mathbf{E}}(t)=
 \begin{bmatrix}
  E_{1}(t) \\
  E_{2}(t)
 \end{bmatrix}.
\end{equation}
The corresponding instantaneous Stokes vector is a 4-element real vector
\begin{equation}\label{eq:vecstok}
\vec{\mathbf{S}}(t)=
 \begin{bmatrix}
  S_{0}(t) \\
  S_{1}(t) \\
  S_{2}(t) \\
  S_{3}(t)
 \end{bmatrix}
\end{equation}
whose elements are the instantaneous Stokes parameters, defined in terms of the complex electric field components as \cite{brosseau98}:
\begin{gather}
S_{0}(t)=E_{1}(t)E_{1}^*(t)+E_{2}(t)E_{2}^*(t), \\ \label{eq:stokesfirst}
S_{1}(t)=E_{1}(t)E_{1}^*(t)-E_{2}(t)E_{2}^*(t), \\
S_{2}(t)=E_{1}(t)E_{2}^*(t)+E_{1}^*(t)E_{2}(t), \\
S_{3}(t)=i\left[E_{1}(t)E_{2}^*(t)-E_{1}^*(t)E_{2}(t)\right]. \label{eq:stokeslast}
\end{gather}
The first component $S_{0}(t)=\vec{\mathbf{E}}(t)^\dagger
\vec{\mathbf{E}}(t)$ is the instantaneous intensity of the field, with
$^\dagger$ denoting the Hermitian conjugate.

The instantaneous Stokes vector completely characterizes the state of
polarization (SOP) of a partially polarized light beam, except its absolute
phase. The conventional Stokes parameters are the ensemble averages of
the instantaneous Stokes parameters. Assuming stationarity and
ergodicity, the conventional Stokes vector is :
\begin{equation}
\vec{\mathbf{S}}=\langle\vec{\mathbf{S}}(t)\rangle=
 \begin{bmatrix}
  \langle S_{0}(t) \rangle \\
  \langle S_{1}(t) \rangle \\
  \langle S_{2}(t) \rangle \\
  \langle S_{3}(t) \rangle
 \end{bmatrix},
\end{equation}
which involves the following time average:
\begin{equation}
\langle X(t) \rangle=\lim_{T \to \infty} {1 \over T} \int_0^T X(t)\mathrm{d}t .
\end{equation}
It is worth to recall that the conventional Stokes parameters are
defined in this way simply because in most experimental setups the
fluctuations of the electric field are produced at optical
frequencies, which are obviously many orders of magnitude higher than
those achievable by the fastest photodetector (so far, ultrafast optical
measurements can only be performed by specific techniques like
nonlinear optical gating or interferometric detection, which are able
to indirectly measure light intensity as well as its time
delay). However, in the next section it is shown that the
instantaneous Stokes vector is a very useful way to characterize
DFDP sources.

\section{Dual-frequency dual-polarization source}\label{sec:dfdp}

\subsection{General equations}
Assuming a dual-frequency dual-polarization (DFDP) source whose two orthogonal modes propagate along the $z$ axis, its transversal electric field can be expressed as:
\begin{equation}
\vec{\mathbf{E}}(t)=
 {E_0 \over \sqrt{2}}e^{-i2\pi\nu t}\left(\begin{bmatrix}a_{1}\\b_1\end{bmatrix}
 +\sqrt{\gamma}e^{-i2\pi\Delta\nu t}\begin{bmatrix}a_2\\b_{2}\end{bmatrix}\right),
\end{equation}
where $\Delta\nu$ is the frequency shift between both modes, and
$\gamma$ accounts for a possible intensity unbalancing between them. Taking into
account the orthogonality condition between the SOP's of the two
polarization modes, the components $a_i$ and $b_i$ in the cartesian
basis can be parameterized in the following general form:
\begin{equation}\label{eq:basemode}
 \begin{gathered}
  a_1=\cos\alpha\cos\epsilon-i\sin\alpha\sin\epsilon,\\
  b_1=\sin\alpha\cos\epsilon+i\cos\alpha\sin\epsilon,
 \end{gathered}
\end{equation}
and
\begin{equation}
 \begin{gathered}
  a_2=-\sin\alpha\cos\epsilon+i\cos\alpha\sin\epsilon,\\
  b_2=\cos\alpha\cos\epsilon+i\sin\alpha\sin\epsilon,
 \end{gathered}
\end{equation}
where $\alpha$ is the polarization ellipse azimuth and $\epsilon$ is its ellipticity \cite{brosseau98}. These equations verify the polarization orthogonality condition $a_1a_2^*+b_1b_2^*=0$.
We set $\alpha=0$ without loss of generality, as we later consider optical elements with arbitrary azimuth. From Eqs.~(\ref{eq:vecstok}-\ref{eq:stokeslast}), the instantaneous Stokes vector of a general DFDP source is:
\begin{equation}
\vec{\mathbf{S}}(t)=I_{0}
 \begin{bmatrix}
  1 \\
  {1-\gamma \over 1+\gamma}\cos(2\epsilon)-2{\sqrt{\gamma} \over 1+\gamma}\sin(2\epsilon)\sin(\Delta\omega t) \\
  2{\sqrt{\gamma} \over 1+\gamma}\cos(\Delta\omega t)\\
  {1- \gamma \over 1+\gamma}\sin(2\epsilon) +2{\sqrt{\gamma} \over 1+\gamma}\cos(2\epsilon)\sin(\Delta\omega t)
 \end{bmatrix},
\end{equation}
where $I_0=|\vec{\mathbf{E}}(t)|^2=(1+\gamma)E_0^2 /2$ and with
$\Delta\omega=2\pi\Delta\nu$ denoting the angular frequency that
corresponds to the interference between both modes. The frequency
shift $\Delta\nu$ can be tuned to values within the radio-frequency
(RF) range, typically from several MHz up to tens of GHz. It can be
observed that the instantaneous intensity $S_0(t)$ of this DFDP
illumination is constant and equal to $I_0$.

\subsection{Linear and circular DFDP source}
\begin{figure}[htbp]
\centerline{\includegraphics[width=8cm]{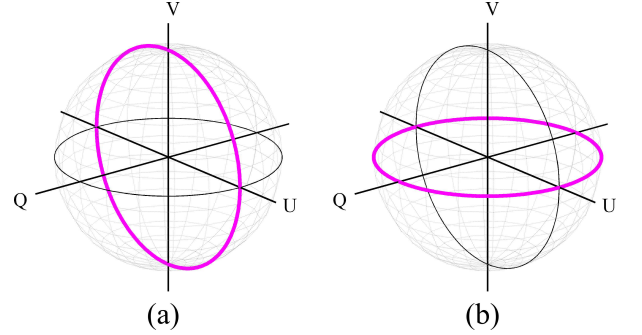}}
\caption{Poincar\'e sphere representation of the instantaneous Stokes vector at the output of the (a) linear DFDP (dual-frequency dual-polarization) balanced source, and (b) circular DFDP balanced source.}
\label{figpoincare}
\end{figure}

On the one hand, if the DFDP source provides two purely-linear orthogonal SOP's, then $\epsilon=0$ and the instantaneous Stokes vector is:
\begin{equation}\label{eq:nonbalancedlinear}
\vec{\mathbf{S}}\mathbf{_{L}}(t)=I_{0}
 \begin{bmatrix}
  1 \\
  {1-\gamma \over 1+\gamma} \\
  2{\sqrt{\gamma} \over 1+\gamma}\cos(\Delta\omega t)\\
 2{\sqrt{\gamma} \over 1+\gamma}\sin(\Delta\omega t) 
 \end{bmatrix}.
\end{equation}
Moreover, if a perfectly balanced source is used ($\gamma=1$), $\vec{\mathbf{S}}\mathbf{_{L}}(t)$ simplifies to:
\begin{equation}
\vec{\mathbf{S}}\mathbf{_{L}}(t)=I_{0}
 \begin{bmatrix}
  1 \\
  0 \\
  \cos(\Delta\omega t)\\
  \sin(\Delta\omega t) 
 \end{bmatrix},
\end{equation}
which corresponds to an instantaneous Stokes vector continuously oscillating at an angular frequency $\Delta\omega$ from a linear $\pm45^\circ$ SOP to a purely circular one, as shown in the Poincar\'e sphere representation included in Fig.~\ref{figpoincare}.a.

On the other hand, if we consider a circular DFDP source, $\epsilon=\pi/4$ and the instantaneous Stokes vector is given by:
\begin{equation}\label{eq:nonbalancedcircular}
\vec{\mathbf{S}}\mathbf{_{C}}(t)=I_{0}
 \begin{bmatrix}
  1 \\
  -2{\sqrt{\gamma} \over 1+\gamma}\sin(\Delta\omega t) \\
  2{\sqrt{\gamma} \over 1+\gamma}\cos(\Delta\omega t)\\
  {1- \gamma \over 1+\gamma} 
 \end{bmatrix},
\end{equation}
which reduces to the following expression for a balanced source:
\begin{equation}
\vec{\mathbf{S}}\mathbf{_{C}}(t)=I_{0}
 \begin{bmatrix}
  1 \\
  -\sin(\Delta\omega t) \\
  \cos(\Delta\omega t)\\
  0
 \end{bmatrix}. 
\end{equation}
It can be observed that the latter instantaneous Stokes vector oscillates along all the possible linear SOP's, as depicted in Fig.~\ref{figpoincare}.b. The advantages of each of these sources for polarimetric measurements are analyzed in the next sections.

\section{Free-space orthogonality breaking sensing}\label{sec:freespaceob}
In this section, we focus on free-space orthogonality breaking sensing, in which the source beam directly impinges on the sample. According to the Mueller calculus, the output Stokes vector $\vec{\mathbf{S}}\mathbf{_{out}}(t)$ after interaction with an anisotropic depolarizing medium is given by:
\begin{equation}
\vec{\mathbf{S}}\mathbf{_{out}}(t)=\mathbf{M}\vec{\mathbf{S}}\mathbf{_{in}}(t),
\end{equation}
where $\mathbf{M}$ is the Mueller matrix of the sample and $\vec{\mathbf{S}}\mathbf{_{in}}(t)$ is the input Stokes vector. It is worth
recalling that this equation is valid for both the instantaneous and
the conventional Stokes vector. All the subsequent results assume
that the propagation direction is kept constant, so the measurement is
made in transmission. The procedure can be equivalently developed for
the reverse direction (reflection or backscattering configuration) provided that
$\mathbf{M}$ is replaced by its corresponding Mueller matrix for the
reverse direction:
\begin{equation}
\hat{\mathbf{M}}=\mathbf{O} \mathbf{M}^{\mathrm{T}} \mathbf{O}^{-1},
\end{equation}
where $\mathbf{O}=\mathrm{diag}(1,1,-1,1)$ if the reversal coordinate
system $\hat{x}\hat{y}\hat{z}$ is set as $\hat{x}=-x$, $\hat{y}=y$ and
$\hat{z}=-z$ \cite{brosseau98} (or alternatively $\hat{x}=x$, $\hat{y}=-y$ and
$\hat{z}=-z$). 

In the remainder of this paper we will focus on the
study of the detected intensity signal $I_{out}(t)$, which is a linear
combination of the input instantaneous Stokes vector determined by the
first row of the Mueller matrix:
\begin{equation}\label{eq:Igeneric}
I_{out}(t)=
 \begin{bmatrix}  
 \mathrm{M_{11}} & \mathrm{M_{12}} & \mathrm{M_{13}} & \mathrm{M_{14}}
 \end{bmatrix}
\vec{\mathbf{S}}\mathbf{_{in}}(t).
\end{equation}

\subsection{Isotropic absorber}
The Mueller matrix of an isotropic absorber is the identity matrix
weighted by the isotropic absorption coefficient $\rho$ such that
$\mathbf{M_{abs}}=\rho\mathbf{I}$, and hence the output intensity is
simply $ I_{out}(t)=\rho I_0$. In this trivial case, it is
straighforward to verify that the orthogonality between the two SOP's
generated by a general DFDP source is unaltered by the sample, and
consequently the measured intensity remains constant in time.

\subsection{Elliptical retarder}
A sample with elliptical birefringence presents a Mueller matrix of the following form \cite{lu96}
\begin{equation}
\mathbf{M_{ER}}=
\begin{bmatrix}
 1 & \vec{0}^{\mathrm{T}} \\
 \vec{0} & \mathbf{M}_{3\times3}
\end{bmatrix},
\end{equation}
where $\mathbf{M}_{3\times3}$ is the $3\times3$ retardance sub-matrix, and $\vec{0}=[0\ 0\ 0]^T$. It can be readily observed that the beam intensity is unaltered by such a type of sample:
\begin{equation}
 I_{out}(t)=I_0,
\end{equation}
as expected for unitary polarization elements \cite{brosseau98}. As a result, the polarimetric orthogonality between the two SOP's provided by the DFDP source is preserved during propagation through birefringent samples. This is the property that was originally used to make orthogonality breaking measurements insensitive to propagation through fibers.

\subsection{Diagonal depolarizer}\label{sec:diagdepol}
The general expression of the Mueller matrix of a diagonal depolarizer sample is \cite{brosseau98}:
\begin{equation}\label{eq:diagdepol}
\mathbf{M_{\Delta}} = \begin{bmatrix}
 1 & 0 & 0 & 0 \\
 0 & P_{L1} & 0 & 0 \\
 0 & 0 & P_{L2} & 0 \\
 0 & 0 & 0 & P_{C}
\end{bmatrix}.
\end{equation}
Diagonal depolarizers are thus defined by the three depolarization
parameters $P_{L1}$, $P_{L2}$, and $P_{C}$. Such depolarizers are the most usual ones. 
In the particular situation of isotropic
linear depolarization (like the one produced by a turbid medium with
randomly-located nearly spherical particles \cite{mishchenko00})
$P_{L1}=P_{L2}$. Furthermore, if completely homogeneous depolarization
is assumed, the sample is usually called a pure depolarizer, and it
can be quantified by a single parameter as $P_{L1}=P_{L2}=P_{C}$.

In any case, the elements of the first row of $\mathbf{M_{\Delta}}$
satisfy $\mathrm{M_{1j}}=0$ for $j\neq1$, so according to Eq.~(\ref{eq:Igeneric}) such
a sample does not modify the orthogonality of the two
orthogonally-polarized SOP's, and the beam intensity is expected to
remain constant in time as $ I_{out}(t)=I_0$.

As a result, the instantaneous Stokes-Mueller calculus detailed in
this article indicates that the orthogonality-breaking sensing
principle is not able to provide a measurement of diagonal
depolarization. This is in contradiction with what was claimed in an
anterior work \cite{fade12}, and with some experimental results obtained in the
same reference which actually seemed to corroborate the possibility of
characterizing depolarization. This aspect is
analyzed and discussed in detail in Section~\ref{sec:discussion}.

\subsection{Diattenuator}

\subsubsection{Linear diattenuator}
The Mueller matrix of a sample showing linear dichroism is:
\begin{widetext}
\begin{equation}\label{eq:mld}
\mathbf{M_{LD}} = \rho
 \begin{bmatrix}
   1 & d\cos(2\phi) & d\sin(2\phi) & 0 \\
   d\cos(2\phi) & {1+T \over 2}+{1-T \over 2}\cos(4\phi) & {1-T \over 2}\sin(4\phi) & 0 \\
   d\sin(2\phi) & {1-T \over 2}\sin(4\phi) & {1+T \over 2}-{1-T \over 2}\cos(4\phi) & 0 \\
   0 & 0 & 0 & T
 \end{bmatrix},
\end{equation}
\end{widetext}
where $\rho=(T_{max}+T_{min})/2$ accounts for the isotropic
absorption, $d=(T_{max}-T_{min})/(T_{max}+T_{min})$ is the
diattenuation coefficient, and
$T=2\bigl[T_{max}T_{min}\bigr]^{\frac{1}{2}}/(T_{max}+T_{min})$. Parameters $T_{max}$ and
$T_{min}\leq T_{max}$ are the maximum and minimum transmittances
respectively, being both of them bounded between 0 and 1, so
$0\leqslant d\leqslant1 $. Finally, the parameter $\phi$ is the linear
dichroism angle, i.e.,~the azimuth of the maximum transmittance
axis. We note that $\mathbf{M_{LD}}$ can be obtained from the Mueller matrix of 
an elliptical diattenuator (explicitly derived in Appendix~\ref{anex:muellerellip} for the sake of generality) 
by setting $\epsilon=0$. An ideal polarizer corresponds to a perfect linear dichroic
sample showing $T_{max}=1$ and $T_{min}=0$, so $\rho=1/2$, $d=1$ and $T=0$.

In this section and in the remainder of this article we will consider
a perfectly balanced source ($\gamma=1$). Indeed, the equations
of the output intensity for a linear diattenuator using a non-balanced
linear and circular DFDP source have been included in Appendix~\ref{anex:nonbalanced} for
the sake of completeness, and they show that the balanced
configuration is actually the most advantageous one for characterizing
the sample properties with the highest dynamics.

According to Eq.~(\ref{eq:Igeneric}), if a linear diattenuator is illuminated with a
balanced linear DFDP source, the first element of the instantaneous
output Stokes vector is:
\begin{equation}\label{eq:Idelnudichro}
I_{outL}(t) = \rho I_0 \left[1 + d\sin(2\phi)\cos(\Delta\omega t) \right],
\end{equation}
where the subscript $L$ indicates the use of linear illumination
states. This equation shows the essential characteristic of the
orthogonality breaking sensing principle, namely an AC component in
the output intensity which is due to the interference of the two
SOP's, partially projected onto each other by the interaction with the
diattenuator. Provided that the source can be tuned to set the
frequency difference to a value that lies within the bandwidth of
commercially-available detectors, it is then perfectly possible to
observe the intensity beatnote with a fast photodiode.

According to the previous equation, the DC and AC components of the output intensity are respectively:
\begin{gather}
I_{outL}^{0} = \rho I_0, \label{eq:I0cirsource} \\
I_{outL}^{\Delta\omega X} = \rho I_0 d\sin(2\phi), \label{eq:IXlinsource}
\end{gather}
the superscript $^X$ accounting for the in-phase component of the beatnote signal at $\Delta\omega$.

The Orthogonality Breaking Contrast (OBC) is a scalar parameter defined from the DC and AC components of the detected signal as:
\begin{equation}\label{eq:obc}
\mathrm{OBC}={\overline{I_{out}^{\Delta\omega}} \over I_{out}^{0}},
\end{equation}
where $\overline{I_{out}^{\Delta\omega}}= \sqrt{\left(I_{out}^{\Delta\omega X}\right)^2 + \left(I_{out}^{\Delta\omega Y}\right)^2 }$ is the amplitude of the detected beatnote signal. In this case, the quadrature component $I_{outL}^{\Delta\omega Y}$ is null, so $\overline{I_{outL}^{\Delta\omega}}= \vert I_{outL}^{\Delta\omega X} \vert$ and the OBC is thus:
\begin{equation}\label{eq:obclinsource}
\mathrm{OBC}_L=d \vert \sin(2\phi) \vert.
\end{equation}
Concerning the phase of the AC signal, in this case it is obviously zero as $\angle{I_{outL}^{\Delta\omega}}= \arctan\left( I_{outL}^{\Delta\omega Y} / I_{outL}^{\Delta\omega X} \right)$. Consequently, when a linear DFDP source is used, the beatnote component does not undergo any phase delay while interacting with the dichroic sample, and the beatnote intensity depends on both the diattenuation coefficient $d$ and the linear dichroism angle $\phi$. It can be observed that the OBC takes a maximum value of $d$ for a linear diattenuator oriented at $\phi=45^{\circ}$. This property has been used in previous works to calibrate the measurement system~\cite{fade12,schaub14}.

If a circular DFDP source is now considered, the different components of the output intensity are:
\begin{gather}
I_{outC}^{0} = \rho I_0, \\
I_{outC}^{\Delta\omega X} = \rho I_0 d \sin(2\phi), \label{eq:IXcirsource} \\
I_{outC}^{\Delta\omega Y} = -\rho I_0 d \cos(2\phi), \label{eq:IYcirsource}
\end{gather}
where the subscript $C$ denotes circular illumination states. In this
case the OBC and the beatnote signal phase are respectively:
\begin{gather}
\mathrm{OBC}_C=d, \label{eq:obccirsource} \\
\angle{I_{outC}^{\Delta\omega}} = 2\phi-\pi/2. \label{eq:phasecirsource}
\end{gather}
From the latter equation, the linear dichroism orientation can be readily obtained by:
\begin{equation}\label{eq:ldangle}
\phi = {1 \over 2} \left (\angle{I_{outC}^{\Delta\omega}}+\pi/2 \right).
\end{equation}
As a result, under circular illumination, the amplitude of the beatnote signal is independent of the dichroism orientation, giving access directly and without ambiguity to the sample dichroism. Moreover, the linear dichroism angle can be directly retrieved by the phase measurement. Such a feature is actually quite advantageous for linear dichroism sensing, as has been recently demonstrated in a microscopic imaging set-up~\cite{ortega15}.

\subsubsection{Elliptical/circular diattenuator}
The complete equations of the detected intensity when the sample
presents elliptical dichroism are included in Appendix~\ref{anex:obellip}. For the sake
of conciseness, we shall only consider here the very specific case of
a circular diattenuator whose Mueller matrix is:
\begin{equation}\label{eq:mcd}
\mathbf{M_{CD}} = \rho
 \begin{bmatrix}
 1 & 0 & 0 & d \\
 0 & T & 0 & 0 \\
 0 & 0 & T & 0 \\
 d & 0 & 0 & 1
 \end{bmatrix},
\end{equation}
which is obtained by setting $\epsilon=\pi/4$ in the Mueller matrix of
an elliptical diattenuator (Appendix~\ref{anex:muellerellip}), $T$ and
$d$ still corresponding to their initial definition given after
Eq.~(\ref{eq:mld}). If the sample is illuminated with a linear DFDP
source, the resulting OBC and phase are:
\begin{gather}
\mathrm{OBC}_L=d, \\
\angle{I_{outL}^{\Delta\omega}} = \pi/2.
\end{gather}
These equations show that such a configuration is actually sensitive to
circular dichroism, which is directly characterized by the OBC, while
the phase of the beatnote signal is constant. This latter information 
actually provides additional information about the dichroism properties 
when a linear DFDP source is used, as it is 0 for linear dichroism, 
$\pi/2$ for circular dichroism, and takes intermediate values for 
elliptical dichroism.

However, if a circular DFDP source is used, the OBC completely vanishes:
\begin{equation}
\mathrm{OBC}_C=0, \\
\end{equation}
which means that such a sample does not break the orthogonality between
the circular SOP's. This is due to the fact that they are precisely
the eigenstates of a circular diattenuator. Consequently, there is no
beatnote signal in this case.

\section{Orthogonality breaking sensing through a waveguide}\label{sec:endoscopicob}
In this section, orthogonality breaking sensing through a waveguide is
considered. It is thus assumed that the source beam is delivered
through a waveguide with Mueller matrix $\mathbf{M_{wg1}}$, when it impinges
on the sample still described by its Mueller matrix $\mathbf{M}$, and
is finally collected by a second waveguide with Mueller matrix
$\mathbf{M_{wg2}}$ (which is not necessarily the same as the
illumination one for the sake of generality), so the output
instantaneous Stokes vector is:
\begin{equation}
\vec{\mathbf{S}}\mathbf{_{out}}(t)=\mathbf{M_{wg2}}\mathbf{M}\mathbf{M_{wg1}}\vec{\mathbf{S}}\mathbf{_{in}}(t).
\end{equation}
We define the intermediate Stokes vector
$\vec{\mathbf{S}}\mathbf{_{out^{\prime}}}(t)=\mathbf{M}\mathbf{M_{wg1}}\vec{\mathbf{S}}\mathbf{_{in}}(t)$
as the instantaneous Stokes vector after interaction of the
fiber-guided DFDP beam with the sample. In general, it can be assumed
that an optical waveguide behaves as a retarder \cite{manhas2015} (with possible
isotropic loss, but with no dichroic effects). We thus model it by the Mueller matrix of a
birefringent element. If we denote $I_{out^{\prime}}$ the first
element of the intermediate Stokes vector, one can readily verify that
%\begin{equation}
$I_{out}=I_{out^{\prime}}$,
%\end{equation}
as the Mueller matrix of a unitary optical element does not modify the
beam intensity. A remarkable implication of this fact is that the
collecting fiber does not modify the beatnote signal possibly
produced by the sample, because such an information is exclusively
carried by the intensity. This means that we can focus our analysis on
the effect of the illuminating waveguide, while light collection
can be performed by any non-dichroic optical waveguide.

Concerning the illuminating fiber, its effect is a modification of the
orthogonal SOP's provided by the source, so it obviously has to be
taken into account. The results for isotropic absorbers, elliptical
retarders, and diagonal depolarizers are not discussed in this
section, as the conclusions obtained in the previous section are valid
regardless the SOP's of the DFDP illuminating beam. Moreover, if we
focus on biomedical applications, circular dichroism is extremely
unusual in biological samples \cite{tuchin06,ghosh11}, so we will exclusively analyze
fiber-guided orthogonality breaking sensing of linearly dichroic
samples.

\subsection{Waveguide acting as a circular retarder}
Let us first consider a waveguide acting as a circular retarder, whose Mueller matrix is:
\begin{equation}
\mathbf{M_{CR}} = \begin{bmatrix}
 1 & 0 & 0 & 0 \\
 0 & \cos(2\theta) & \sin(2\theta) & 0 \\
 0 & -\sin(2\theta) & \cos(2\theta) & 0 \\
 0 & 0 & 0 & 1
\end{bmatrix},
\end{equation}
$\theta$ being the optical rotation angle. If such a waveguide is used
to illuminate a linear diattenuator with a DFDP source, it is easily
shown that the results obtained in the previous section hold, up to a
rotation angle of value $\theta$ due to optical rotation in the
waveguide. Typically, one has $\mathrm{OBC}_L=d \vert
\sin(2(\phi+\theta)) \vert$ with a linear DFDP source, whereas
$\mathrm{OBC}_C=d$ and $\angle{I_{outC}^{\Delta\omega}} =
2(\phi+\theta)-\pi/2$ with a circular DFDP source.

From these results, it is interesting to note that with circular DFDP
illumination states, the presence of an illuminating waveguide acting
as a rotator does not prevent from measuring the diattenuation
coefficient, while the phase can be determined up to an additive 
term depending on the fiber.

\subsection{Waveguide acting as a linear retarder}
More typically, an optical waveguide behaves as a linear retarder, whose general Mueller matrix is:
\begin{widetext}
\begin{equation}
\mathbf{M_{LR}} = \begin{bmatrix}
 1 & 0 & 0 & 0 \\
 0 & \cos\delta\sin^{2}(2\psi)+\cos^{2}(2\psi) & (1-\cos\delta)\cos(2\psi)\sin(2\psi) & -\sin\delta\sin(2\psi) \\
 0 & (1-\cos\delta)\cos(2\psi)\sin(2\psi) & \cos\delta\cos^{2}(2\psi)+\sin^{2}(2\psi) & \sin\delta\cos(2\psi) \\
 0 & \sin\delta\sin(2\psi) & -\sin\delta\cos(2\psi) & \cos\delta
\end{bmatrix}, 
\end{equation}
\end{widetext}
with $\psi$ denoting the linear birefringence orientation, whereas $\delta$ stands for the retardation introduced between the birefringence slow and fast axes~\cite{brosseau98}.

Illuminating a linear diattenuator through a linearly birefringent waveguide gives a lengthy expression of the output intensity, that can be simplified to:
\begin{gather}
I_{outL}^{0} = \rho I_0, \\
\begin{split}
  I_{outL}^{\Delta\omega X} & = I_0 \rho d \biggl[\sin (2 \psi ) \cos(2(\phi-\psi)) \\
&\qquad + \cos(2 \psi )\sin(2(\phi-\psi)) \cos \delta \biggr],
\end{split} \\
I_{outL}^{\Delta\omega Y} =  I_0 \rho d \sin (\delta ) \sin (2 (\phi -\psi )).
\end{gather}
It can be checked that when the linear birefringence axis of the
waveguide is parallel to the linear dichroism $(\psi=\phi)$, the previous equations are exactly the same as the
results obtained without the waveguide (Eqs.~(\ref{eq:I0cirsource}-\ref{eq:IXlinsource})), i.e.,~$I_{outL,\delta
  \neq 0}^{\Delta\omega}(t)-I_{outL,\delta=0}^{\Delta\omega}(t)
=0$. However, in general the linear birefringence introduced by
the waveguide can modify the calculated parameters, and produces an
intrincate expression of the OBC. Let us define the bias of the
measurement as
\begin{equation}\label{eq:bias}
\mathrm{B}_L(t)={I_{outL,\delta \neq 0}^{\Delta\omega}(t)-I_{outL,\delta=0}^{\Delta\omega}(t) \over I_{outL}^{0}}.
\end{equation}
With a linear DFDP source, the bias due to a linear birefringent waveguide then reads
\begin{equation}\begin{split}
\mathrm{B}_L (t)&= 2d\sin\left({\delta \over 2}\right) \biggl[ -\cos(2\psi)\sin\left({\delta \over 2}\right)\cos(\Delta\omega t) \\& + \cos\left({\delta \over 2}\right)\sin(\Delta\omega t) \biggr] \sin(2(\phi-\psi)).
\end{split}
\end{equation}
If we consider the particular case of a slightly birefringent fiber
($\delta\ll 1$), the series expansion of $\mathrm{B}_L (t)$ in $\delta$ leads at first
order in $\delta$ to
\begin{equation}
\mathrm{B}_L(t) = d \sin(2(\phi-\psi)) \sin(\Delta\omega t) \delta + O(\delta).
\end{equation}
If a circular DFDP source is now taken into account, the output
intensity components are:
\begin{gather}
I_{outC}^{0} = \rho I_0, \\
I_{outC}^{\Delta\omega X} = I_{outL}^{\Delta\omega X} \label{eq:IxDFDCPwg}\\
\begin{split}
I_{outC}^{\Delta\omega Y} &= - I_0 \rho d \biggl[\cos (2 \psi ) \cos(2(\phi-\psi)) \label{eq:IyDFDCPwg} \\
&\qquad - \sin(2 \psi )\sin(2(\phi-\psi)) \cos \delta \biggr].
\end{split}
\end{gather} 
Again, it can be verified that these equations coincide with the
free-space sensing ones for the particular case in which the fiber
optical axis is parallel to the sample linear dichroism
$(\psi=\phi)$. Nonetheless, any other situation leads to a
modification of the results by the retarding action of the waveguide.
Applying the bias defined in Eq.~(\ref{eq:bias}) to the circular DFDP source using Eqs.~(\ref{eq:IxDFDCPwg}) and (\ref{eq:IyDFDCPwg}), one gets
\begin{equation}
\mathrm{B}_C(t) = 2d\sin\left({\delta \over 2}\right)^2 \sin(2(\phi-\psi)) \cos(\Delta\omega t - 2\psi).
\end{equation}
If a slighly birefringent fiber is again considered, it can be verified that the bias is null at first order in $\delta$, and that 
\begin{equation}
\mathrm{B}_C(t) =d  \sin(2(\phi-\psi)) \cos(\Delta\omega t - 2\psi)\frac{\delta ^2}{2} + O(\delta^2).
\end{equation}
The bias in this case is actually at order 2 in $\delta$. This is an
interesting result, showing that the circular DFDP source is more
advantageous for orthogonality breaking sensing through a linear
birefringent waveguide, since the measured OBC is less prone to be
biased by a slight birefringence in the waveguide.

\section{Discussion}\label{sec:discussion}
In this last section, we complement the preceding
description of the orthogonality breaking signatures on various optically anisotropic
samples by a discussion on the ability of this technique to sense
depolarization. Indeed, even though calculations using
the instantaneous Mueller-Stokes formalism in Section~\ref{sec:freespaceob} show that
the technique yields no orthogonality breaking contrast on a diagonal depolarizer,
previous experimental results~\cite{fade12} seem to be in contradiction with
this statement.

In order to clarify this essential aspect, we first performed a verification measurement 
using a linear DFDP blue source at $\lambda=488$ nm whose development was detailed in a previous work~\cite{schaub14}. 
The measurements were performed in free-space for two different diffusing samples, namely a blue paper and a red paper. The detector was fixed in a reflection configuration, with an incidence angle of roughly 45$^{\circ}$ on the sample. This measurement configuration was kept identical for both samples. As explained in detail in a previous work dealing with spectro-polarimetric imaging of diffuse objects~\cite{alouini2008}, illuminating both samples with a visible blue source yields two different situations. On the one hand, the main contribution to the backscattered light when the blue paper is measured comes from volume multiple scattering, as the incoming light is weakly absorbed. This type of scattering is very depolarizing, and its Mueller matrix corresponds to a diagonal depolarizer. On the other hand, when the red paper is used as a sample, the incident blue light beam is strongly absorbed by the red pigments. Consequently, the fraction of light reflected towards the detector mostly results from surface scattering, which implies much weaker depolarization and possible anisotropy of the reflection coefficients.

\begin{figure}[htpb]
\begin{center}
\includegraphics[width=7cm]{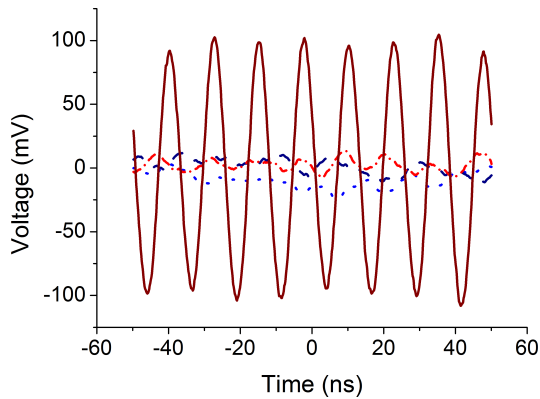}
\caption{AC component of the detected backscattered light under blue illumination for a blue paper and linear modes along the $0^{\circ}-90^{\circ}$ directions (dotted light blue curve) and the $\pm45^{\circ}$ directions (dashed dark blue curve), and for a red paper along the same directions (dashed-dotted light red curve and solid dark red curve respectively). \label{fig:diffuse}}
\end{center}
\end{figure}

The measurements are presented in Fig.~\ref{fig:diffuse}, which shows the AC component of the detected intensity for both samples using the linear DFDP source with the states oriented in two different ways, namely along the $0^{\circ}-90^{\circ}$ and the $\pm45^{\circ}$ directions. In the first configuration, the two linear states of the DFDP beam respectively correspond to the so-called $s$ and $p$ polarization components at the surface, whereas in the second configuration both linear states have partial projections on the $s$ and $p$ directions. It can be observed that the AC component for the blue paper is almost zero regardless of the linear states orientation (dotted light blue curve and dashed dark blue curve respectively). In this case, the dominant effect is volume multiple scattering as mentioned above, which constitutes an initial verification of the fact that diagonal depolarization due to the sample does not produce an orthogonality breaking signal. Regarding the red paper, one can note that there is a slight variation from roughly no AC signal when the linear modes are oriented at $0^{\circ}-90^{\circ}$ (dashed-dotted light red curve), to a substantial AC component when the modes are along $\pm45^{\circ}$ (solid dark red curve). This result is in agreement with the expected behaviour, as in this case surface effects are not masked by volume scattering, and thus slight differences in the reflectance coefficients for the $s$ and $p$ components of the incident light beam can result in dichroic effects in the polarized fraction of the detected backscattered light.

An appropriate interpretation of the aforementioned discrepancy between these 
results and some of our previous ones requires a deeper
insight into the physical origins of depolarization. In the most
general case, this generic notion must
obviously encompass (i) the optical anisotropic properties of the
sample considered; but also (ii) its local structural organization,
including spatial randomization effects; (iii) the properties of the
illumination field (e.g.  spectral bandwidth); and (iv) the
characteristics of the detection setup (e.g. numerical
aperture~\cite{pouget12} and spatial/spectral resolution). Describing
a depolarizing medium with the Mueller matrix of a diagonal
depolarizer (as $\mathbf{M_{\Delta}}$ in Eq.~(\ref{eq:diagdepol}))
implicitly assumes full incoherent averaging over at least one of the above
aspects. However, in a broad range of experimental setups the
characteristics of the sensing/imaging system only lead to a partial
averaging operation. As a result, the description of the sample by a
diagonal depolarizing Mueller matrix is no longer physically
appropriate in such cases.

In the remainder of this section, we propose to use a simple stochastic model
of the Mueller matrix of a depolarizing anisotropic medium consisting
of the action of random linear dichroic elements. It must be noted that this description does not represent a fully comprehensive model of depolarization, as many physical parameters involved in light depolarization are neglected. However, it is shown that it succesfully allows us to characterize the progressive transition from
a deterministic anisotropic sample to a strongly random one
(characterized by a Mueller matrix of a diagonal depolarizer) as the
averaging conditions evolve. We then analyze the polarimetric
properties of the resulting ``macroscopic'' depolarizing Mueller
matrix obtained, and we characterize the orthogonality breaking signatures produced by
such a sample and physical sensing conditions.  A simple experimental
validation using a synthetic depolarizing sample is also
included. These results and discussions finally allow us to validate
the calculations presented in this work, and to understand how the
subtle aspects discussed regarding the measurement conditions were
actually involved in our previous experimental measurements.

\subsection{Stochastic depolarization model of an ensemble of random linear diattenuators}

We consider the specific case of a dichroic depolarizing medium in
which depolarization is due to the heterogeneity of its anisotropy
properties. For that purpose, we consider that the incoming beam urdergoes random 
local dichroic interactions, each polarization
transformation having a diattenuation coefficient $d_\mu$, linear
dichroism angle $\phi_\mu$, isotropic absorption $\rho_\mu$, and
transmission parameter $T_\mu$, where $\mu$ denotes one realization of a random
event. The effect of a given random event on a DFDP illumination beam
is obviously strictly equivalent to the one studied in
Section~\ref{sec:freespaceob}, and can consequently be represented by
a single Mueller matrix $\mathbf{M_{LD}}_\mu$ with the same form as
that given in Eq.~(\ref{eq:mld}). Consequently, the Mueller matrix of
a single random event is not depolarizing (Mueller-Jones matrix), as
the individual polarization transformation is purely deterministic. In
this case, there is a well-known one-to-one relationship between
$\mathbf{M_{LD}}_\mu$ and its corresponding Jones matrix, as stated in
Eq.~(\ref{eq:muellerjones}) of Appendix A \cite{anderson94,chipman10}.

The macroscopic Mueller matrix of the sample that determines the
detected intensity now implies averaging over random events,
i.e.,~$\mathbf{M_{LD}^\Delta} = \langle \mathbf{M_{LD}}_\mu \rangle
_\mu$. For the sake of simplicity, let us assume that the random
variable $\phi_\mu$ is independent from $d_\mu$ and
$\rho_\mu$. Regarding its probability density function, we propose to
adopt a convenient statistical model for angular random variables,
namely the Wrapped-Gaussian Distribution~(WGD), with average value
$\bar{\phi}$ and variance $\sigma^2_\phi$~\cite{mardia99}. The definition and main
properties of WGD's, which basically correspond to normal probability
distributions `wrapped' around the unit circle, are recalled in
Appendix~\ref{anex:wgd}. Using these properties, the ensemble-averaged
Mueller matrix of the sample is:
\begin{widetext}
\begin{equation}\label{eq:mldrandom}
\mathbf{M_{LD}^\Delta} = \langle \mathbf{M_{LD}}_\mu \rangle _\mu = \bar{\rho}\begin{bmatrix}1 & \bar{d} \cos 2\bar{\phi}\,e^{-2\sigma_\phi^2} & \bar{d} \sin 2\bar{\phi}\,e^{-2\sigma_\phi^2} & 0\\
\bar{d} \cos 2\bar{\phi}\,e^{-2\sigma_\phi^2} & \frac{1+\langle T \rangle}{2}+ \frac{1-\langle T \rangle}{2} \cos 4\bar{\phi}\,e^{-8\sigma_\phi^2} & (1-\langle T \rangle)\cos 2\bar{\phi} \sin 2\bar{\phi}\,e^{-8\sigma_\phi^2} & 0\\
\bar{d} \sin 2\bar{\phi}\,e^{-2\sigma_\phi^2}&(1-\langle T \rangle)\cos 2\bar{\phi} \sin 2\bar{\phi}\,e^{-8\sigma_\phi^2} & \frac{1+\langle T \rangle}{2}- \frac{1-\langle T \rangle}{2} \cos 4\bar{\phi}\,e^{-8\sigma_\phi^2} & 0\\
0 & 0 & 0 & \langle T \rangle \end{bmatrix},
\end{equation}
\end{widetext}
where, for the sake of generality, random variables $T_{min}$ and
$T_{max}$ are simply assumed to admit average values $\overline{T}_{min}$
and $\overline{T}_{max}$, hence
$\bar{\rho}=(\overline{T}_{max}+\overline{T}_{min})/2$, and we set
$\bar{d}=(\overline{T}_{max}-\overline{T}_{min})/2\bar{\rho}$.

From this matrix, the output intensity can be obtained from
Eq.~(\ref{eq:Igeneric}) (in the same way as the calculations detailed
in Sections~\ref{sec:freespaceob} and~\ref{sec:endoscopicob}), and it
is quite straightforwardly shown that the orthogonality breaking
contrast obtained for this sample is
\begin{equation} \label{res_dic_ID_stoch_amp_L}
\mathrm{OBC}_L=\bar{d} e^{-2\sigma_\phi^2} |\sin 2
  \bar{\phi}|
\end{equation}
in the case of linear input polarization states, whereas circular
states would yield
\begin{equation} \label{res_dic_ID_stoch_amp_C}
\mathrm{OBC}_C= \bar{d}  e^{-2\sigma^2_\phi},\quad \text{and }\quad \angle I_{outC}^{\Delta\omega} =2\bar{\phi}-\frac{\pi}{2}.
\end{equation}
These results are very similar to the case of deterministic
transformations of the state of polarization
(Eq.~(\ref{eq:obclinsource}) and Eqs.~(\ref{eq:obccirsource}) and
(\ref{eq:phasecirsource})), up to a ``fading'' factor of the beatnote
amplitude equal to $e^{-2\sigma_\phi^2}$. As a result, a strong
dispersion of the dichroism orientations would blur the orthogonality
breaking signal produced by the diattenuation properties of the
sample. Once again, one can note that using circular input states is
more favourable, since the mean value of the diattenuation orientation
$\bar{\phi}$ can be retrieved from the measurement of the beatnote
phase, provided the beatnote amplitude is not completely attenuated.

To further analyze the previous results, let us now assume that the
diattenuation angle $\phi_\mu$ is the only random parameter, $d$,
$\rho$ and $T$ being now considered as deterministic. On the one hand,
it can be immediately observed that when $\sigma_\phi \to 0$,
$\mathbf{M_{LD}^\Delta}_{\sigma_\phi \to 0}=\mathbf{M_{LD}}$, which
corresponds to the trivial case of a deterministic sample, whose
measurement using the orthogonality breaking technique obviously yields the same results
as those obtained in the previous sections. On the other hand, when
the angular distribution becomes strongly randomized
(i.e.,~$\sigma_\phi \gg 1$), the Mueller matrix tends to the form of a
diagonal depolarizer:
\begin{equation}
\mathbf{M_{LD}^\Delta}_{\sigma_\phi\gg 1}=\rho\ \mathrm{diag}\bigl[1,\, (1+T)/2,\, (1+T)/2,\, T\bigr].
\end{equation}
We recall that if dichroism is perfect ($d=1$), then $T=0$ and the
previous matrix corresponds to a sample that completely depolarizes
the fourth element of the Stokes vector, and reduces by 0.5 the DOP of
any linear input SOP. For other values of $T$, the
depolarization strength of such a diagonal depolarizer for each Stokes
vector element varies. In any case, it is verified that the resulting
intensity is constant $ I_{out}(t)=\rho I_0$ (in agreement with the
results obtained in Section \ref{sec:diagdepol} for diagonal
depolarizers), so no orthogonality breaking signal appears. Apart from
that, we note that $\mathbf{M_{LD}^\Delta}_{\sigma_\phi\gg 1}$ turns
out to be proportional to the identity matrix (isotropic absorption)
when $T\to 1$ (or equivalently $d\to0$).  These features confirm that
the presented approach makes it possible to simply model the
continuous transition from a non-depolarizing sample characterized by
a deterministic polarization transformation on the one hand, to a
fully depolarizing sample depending on the statistical properties of
the random diattenuation parameters on the other hand.

\begin{figure}[ht]
\begin{center}
\includegraphics[width=8cm]{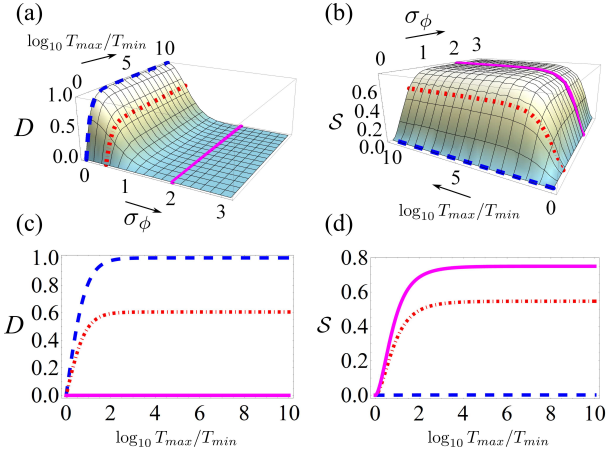}
\caption{(a) Evolution of the effective diattenuation coefficient $D$
  as a function of angular dispersion $\sigma_\phi$ and of $\log_{10}
  T_{max}/T_{min}$. (b) Evolution of the Cloude entropy ${\cal S}$ as
  a function of angular dispersion $\sigma_\phi$ and of $\log_{10}
  T_{max}/T_{min}$. Evolution of the diattenuation coefficient $D$ (c)
  and of the Cloude entropy ${\cal S}$ (d) as a function of $\log_{10}
  T_{max}/T_{min}$ for $\sigma_\phi=0$ (blue), $\sigma_\phi=1/2$
  (red), and $\sigma_\phi=2$ (magenta). \label{fig:entrop}}
\end{center}
\end{figure}

The polarimetric properties of the stochastic Mueller matrix obtained
above can now be quantified by several parameters. The first one is
the diattenuation coefficient, that can be calculated from
$\mathbf{M_{LD}^\Delta}$ (assuming again that $\phi_\mu$ is the only
random parameter) by~\cite{lu96}
\begin{equation}\label{diatt}
   D=\frac{\sqrt{\sum_{j=2}^4 (\mathbf{M_{LD}^\Delta})^2_{1j}}}{\mathbf{(M_{LD}^\Delta})_{11}}=%\frac{ \sqrt{ \sum_{j=2}^4\tr(\bar{D}^{(j)})^2}}{\tr(\bar{D}^{(1)})^2},
   d e^{-2\sigma_\phi^2},
\end{equation}
showing that it scales exactly as the OBC with the angular
dispersion. As a result, on this sample, the orthogonality breaking
technique using circular states for instance would provide a direct
measure of the \emph{effective} linear diattenuation of the sample (as
$\mathrm{OBC}_C=D=d e^{-2\sigma_\phi^2}$), and of the average
diattenuation orientation $\bar{\phi}$. The evolution of the effective
dichroism $D$ is plotted in Fig.~\ref{fig:entrop}.(a) as a function of
$\sigma_\phi$ and of $\log_{10} T_{max}/T_{min}$ (which is 0 for an
isotropic sample and tends to infinity for a perfect polarizer). It
can be seen that the effective linear dichroism rapidly decreases with
$\sigma_\phi$, whereas it increases for higher $\log_{10}
T_{max}/T_{min}$ as expected. The evolution of $D$ for $\sigma_\phi=0
$ corresponds to the diattenuation coefficient of a linear
diattenuator with a fixed deterministic orientation.

It is now interesting to analyze the depolarizing properties of the $\mathbf{M_{LD}^\Delta}$ matrix. There are several depolarization metrics that can be used to
quantify the depolarizing properties of a
sample~\cite{lu96,cloude86,gil85,chipman05,ossikovski10,ortega15b}. In this work
we use the Cloude entropy, which is a well-established metric to
characterize the overall depolarizing nature of a given Mueller
matrix~\cite{cloude86}.  The Cloude entropy is given by $ {\cal
  S}=-\sum_{i=1}^4 \lambda'_i \log_4 \lambda'_i$, where
$\lambda'_i=\lambda_i/\sum_{j=1}^4\lambda_j$ are the normalized
eigenvalues of the $4\times 4$ Cloude coherency
matrix~\cite{cloude86}, whose derivation from $\mathbf{M_{LD}^\Delta}$
is detailed in Appendix~\ref{anex:cloude}. The Cloude entropy ${\cal
  S}$ is plotted in Fig.~\ref{fig:entrop}.(b) as a function of
$\sigma_\phi$ and of $\log_{10}T_{max}/T_{min}$. To facilitate the
physical interpretation, the evolution of parameters $D$ and ${\cal
  S}$ with $\log_{10} T_{max}/T_{min}$ for three different values of
$\sigma_\phi$ (namely 0, 1/2 and 2) are respectively plotted in
Figs.~\ref{fig:entrop}.(c) and (d).

It can be seen in Fig.~\ref{fig:entrop}.(b) that the Cloude entropy $
{\cal S}$ increases with $\sigma_\phi$, thus evidencing that
depolarization is stronger as the angular dispersion grows. On the
other hand, the Cloude entropy increases with $\log_{10}
T_{max}/T_{min}$.  The Mueller matrix of the sample at $\sigma_\phi=0$
or $\log_{10} T_{max}/T_{min}=0$ corresponds to a deterministic
Mueller matrix, hence leading to a null entropy. The Cloude entropy
reaches a maximum at $ {\cal S}=0.75$ for high values of $\sigma_\phi$
and significant anisotropy ($\log_{10}T_{max}/T_{min}\neq 0$).  It can
be noted that the maximum Cloude entropy does not reach unity, simply
because the stochastic model of the sample considered does not lead to
a complete depolarization of any input SOP, as we have only considered
the subset of random linearly dichroic events without including
elliptical dichroism.  

More generally, the joint analysis of the plots in
Fig.~\ref{fig:entrop}.(a) and \ref{fig:entrop}.(b) clearly confirms
the gradual evolution of $\mathbf{M_{LD}^\Delta}$ from a deterministic
Mueller matrix of a diattenuator (${\cal S}=0$, $D\neq 0$) to a
depolarizing Mueller matrix (maximum ${\cal S}$, $D\to 0$) when the
angular dispersion $\sigma_\phi$ increases, as predicted by
Eq.~(\ref{eq:mldrandom}). In other words, we observe that the
intrinsic dichroic properties of the sample gradually vanishe as more
orientations of the dichroism are taken into account by increasing
$\sigma_\phi$, providing the sample with a ``macroscopic''
depolarizing nature.

\subsection{Interpretation of experimental results}

\begin{figure}[ht]
\begin{center}
  \includegraphics[width=7cm]{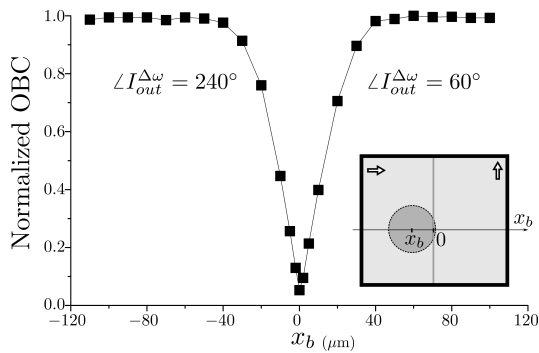}
  \caption{Evolution of the OBC and phase as a function of
    beam position. Inset: sketch of the synthetic sample composed of a
    juxtaposition of two orthogonally oriented polarizing sheets. The
    illumination beam centered in $x_b$ is made to horizontally scan
    across the sample, $x_b=0$ corresponding to the juxtaposition edge
    of the two polarizing sheets. \label{fig:polar}}
\end{center}
\end{figure}

The previous results can be easily confirmed on a simple laboratory
experiment. For that purpose, we used the DFDP visible source
($\lambda=488$ nm) emitting linear polarization states to shine a sample 
composed of two orthogonally-oriented linear
polaroid sheets placed in juxtaposition to each other. A sketch of the
sample is presented in the inset of Fig.~\ref{fig:polar}. The position
of the laser beam is then displaced along the sample, whose Mueller
matrix $\mathbf{M_s}$ can be written
\begin{equation}
 \mathbf{M_s} = A\ \mathbf{M_{LD}}_{\phi=0} + (1-A)\mathbf{M_{LD}}_{\phi=\pi/2},
\end{equation}
where $A$ is the fraction of the laser spot area lying on the
horizontal polarizer. In the central position $A=1/2$, so both
polarizers equally contribute to the detected intensity, and the
resulting Mueller matrix is:
\begin{equation}
 \mathbf{M_s}_{center} = \rho\ \mathrm{diag}\bigl[1,\, 1,\, T,\, T\bigr].
\end{equation}
It is readily observed that such a Mueller matrix corresponds to a
diagonal depolarizer. The commercial polaroid sheets used satisfy $
T\simeq 0$, so when the laser spot is centered in the middle of the
synthetic sample proposed in this section (i.e.,~$x_b=0$), it behaves
as a diagonal depolarizer that completely depolarizes the third and
fourth elements of the input Stokes vector, without altering the
second one appart from the isotropic absorption. As a result, such
a sample provides an OBC that evolves from a maximum value when the spot
entirely lies on a single polaroid sheet, to a null value when it is
placed at the center. The same conclusion can be reached by separately
calculating the output intensities for $\phi=0$ and $\phi=\pi/2$ using
Eq.~(\ref{eq:IXlinsource}) and adding them, which results in a
destructive interference between both beatnote signals in the central
position since they show a relative phase of $\pi$.

The OBC for this sample was measured using the previously reported
setup in transmission \cite{ortega15}. The evolution of the OBC as a
function of the beam position is plotted in Fig.~\ref{fig:polar}. This
experimental curve follows the expected behaviour, evidencing how the
maximum orthogonality breaking contrast obtained on a perfectly
dichroic sample (extremal positions of the sample) is gradually lost
as the beam simultaneously interacts with two orthogonally-oriented
polarizers. Indeed, it is confirmed that when the beam is centered,
both contributions from each half of the beam destructively interfere
on the detector, resulting in a null beatnote amplitude.

In light of all the above results, we are now able to provide an
interpretation of the discrepancy between the theoretical predictions
presented in Section~\ref{sec:freespaceob} and the experimental
observations at $\lambda=1.55$ $\mu$m reported in~\cite{fade12}. In
that work, orthogonality breaking measurements were carried out in a
fibered configuration, and then compared to the control values
determined with a standard free-space Stokes polarimeter at
$\lambda=1.55$ $\mu$m. According to the discussions presented in this
section, the difference in the experimental conditions of both
measurements can have strong consequences on the measured
depolarization. Indeed, the standard Stokes polarimetry measurements
were carried out in free-space, with a relatively high spot size on
the sample and high numerical aperture for the light collection. Such an
experimental configuration effectively implies a spatial averaging
operation over the sample surface and over several spatial coherence
areas (speckle grains) as discussed in~\cite{pouget12}. However, the
OB signals detected on the same samples were collected through a
standard single-mode SMF28 optical fiber whose FC-APC connector end
was placed in vicinity of the samples. Under such conditions, the spatial or angular averaging is
very moderate, which presumably corresponds to an intermediate position
in the aforementioned transition from a non-depolarizing sample to a
depolarizing one. In that case, both the diattenuation coefficient and
the sample depolarization lie between their respective maximum and
minimum values, as shown in Fig.~\ref{fig:entrop}. Therefore, in light
of the thorough modelization proposed in this work and of the
discussion detailed in this section, we can conclude that the OBC
measured in~\cite{fade12} were most likely due to a moderate spatial
averaging of the local diattenuation properties of the samples rather
than pure depolarization.

\section{Conclusion}\label{sec:concl}

In this work, the instantaneous Stokes-Mueller formalism has been applied to conveniently describe the DFDP beam used in the polarimetric sensing by orthogonality breaking technique, and to model the interaction of such a beam with anisotropic depolarizing media. Based on this formalism, we have thoroughly analyzed the characteristics of the orthogonality breaking signal after interacting with birefringent, dichroic, and depolarizing samples. It has thus be confirmed that this measurement technique provides a direct characterization of dichroic samples, and that using a circular DFDP source makes it possible to readily determine the diattenuation magnitude and orientation from the amplitude and the phase of the detected intensity. Moreover, the insensitivity of this technique to birefringence has been confirmed both theoretically and experimentally. This constitutes an interesting property for remote polarimetric measurements, especially for endoscopic applications involving optical fibers. We have consequently characterized the influence of a birefringent waveguide on the detected beatnote component, showing that the orthogonality breaking contrast is affected by a slight bias at order two in the residual
fiber birefringence when a circular DFDP source is used.

Lastly, we have proposed a simple stochastic model of a depolarizing sample composed of randomly-oriented linear diattenuators. Such an analytical model, along with the results of a simple and meaningful laboratory experiment, clearly illustrates the gradual orthogonality breaking contrast vanishing as the orientation randomization of the sample increases, due to the destructive interference of the dephased individual orthogonality breaking beatnotes. All these considerations have led us to reinterpret our first experimental results, in which the observed orthogonality breaking contrast was most likely due to the effective diattenuation of the samples rather than to their depolarizing properties. Consequently, in light of the comprehensive model and the experimental measurements presented in this work, it is concluded that the orthogonality breaking technique is insensitive to diagonal depolarization. This is an important property to be highlighted. Indeed, most polarimetric techniques are influenced by depolarization, which encompasses many physical aspects, including the structural properties of the sample, the detection geometry, and the source and detector bandwidth. Being exclusively sensitive to dichroism, the orthogonality breaking technique is thus remarkably advantageous for characterizing such a parameter without being affected by other sample properties, hence potentially leading to a more specific and robust sample characterization.

The general method presented in this work provides an in-depth
analysis on the physical origin of the detected signals in different
measurement configurations. These results allow orthogonality breaking
sensing to be adequately modeled, which paves the way for the optimal
design of orthogonality breaking imaging systems with the capacity to perform direct and
fast polarimetric measurements at high dynamics. The future
development of this technique includes a systematic comparative study
with standard polarimetric imaging techniques in various application
contexts, and its extension to remote endoscopic measurements
through fiber bundles for biomedical applications.

\bigskip

\appendix
\renewcommand{\theequation}{\thesection-\arabic{equation}}

\section{}\label{anex:muellerellip}
A characteristic property of the Jones matrix of a general dichroic sample is that its two eigenvalues are real and take the form $\lambda_1=T_{max}^{1/2}$ and $\lambda_2=T_{min}^{1/2}$, where $T_{max}$ and $T_{min}$ are respectively the maximum and minimum transmittances. If we consider the eigenvector $\vec{\mathbf{E}}\mathbf{_{eig}}$ corresponding to the greatest eigenvalue:
\begin{equation}
\vec{\mathbf{E}}\mathbf{_{eig}}=
 \begin{bmatrix}
  a \\
  b
 \end{bmatrix},
\end{equation}
the Jones matrix of any elliptic dichroic sample is, according to Ref.~\cite{kliger90}:
\begin{equation}
\mathbf{J_{ED}}=
 \begin{bmatrix}
  \lambda_1 a a^*+\lambda_2 b b^* & \left( \lambda_1-\lambda_2 \right) ab^* \\
  \left( \lambda_1-\lambda_2 \right) ba^* & \lambda_2 a a^*+\lambda_1 b b^*
 \end{bmatrix}.
\end{equation}
If $a$ and $b$ are parameterized according to the general expressions given in Eq.~(\ref{eq:basemode}), the Mueller matrix $\mathbf{M_{ED}}$ of an elliptical diattenuator can be readily obtained by applying the well-known relationship between a Jones matrix and its equivalent Mueller-Jones matrix:
\begin{equation} \label{eq:muellerjones}
 \mathbf{M}=\boldsymbol{\mathrm{T}}\bigl(\mathbf{J} \otimes \mathbf{J}^*\bigr)\boldsymbol{\mathrm{T}}^{-1},
\end{equation}
where $\otimes$ stands for Kronecker product and matrix $\boldsymbol{\mathrm{T}}$ is 
\begin{equation} \label{eq:matrixt}
 \boldsymbol{\mathrm{T}}=\begin{pmatrix}1&0&0&1\\1&0&0&-1\\0&1&1&0\\0&i&-i&0\end{pmatrix},
\end{equation}
which satisfies $\boldsymbol{\mathrm{T}}^{-1}= 1/2 \boldsymbol{\mathrm{T}}^{\dagger}$ \cite{anderson94}. From the previous equations, the explicit form of $\mathbf{M_{ED}}$ is found to be:
\begin{widetext}
\begin{equation}
\mathbf{M_{ED}} = \rho
 \begin{bmatrix}
 1 & dC_{2\phi}C_{2\epsilon} & dS_{2\phi}C_{2\epsilon} & dS_{2\epsilon} \\
 dC_{2\phi}C_{2\epsilon} & {1+3T \over 4}+{1-T \over 4} \left[ C_{4\epsilon}+2C_{4\phi}S^{2}_{2\epsilon} \right] & {1-T \over 2}S_{4\phi}C^{2}_{2\epsilon} & {1-T \over 2}C_{2\phi}S_{4\epsilon} \\
 dS_{2\phi}C_{2\epsilon} & {1-T \over 2}S_{4\phi}C^{2}_{2\epsilon} & {1+3T \over 4}+{1-T \over 4} \left[ C_{4\epsilon}-2C_{4\phi}S^{2}_{2\epsilon} \right] & {1-T \over 2}S_{2\phi}S_{4\epsilon} \\
 dS_{2\epsilon} & {1-T \over 2}C_{2\phi}S_{4\epsilon} & {1-T \over 2}S_{2\phi}S_{4\epsilon} & {1+T \over 2}-{1-T \over 2}C_{4\epsilon}
\end{bmatrix},
\end{equation}
\end{widetext}
where the compact notation $C_{k\phi}^n=\cos^n(k\phi)$ and $S_{k\phi}^n=\sin^n(k\phi)$ has been used. The well-known matrix of an ideal elliptical diattenuator~\cite{chipman10} results from substituting $T_{max}=1$ and $T_{min}=0$, so $\rho=1/2$, $d=1$ and $T=0$ in the previous equation.

\section{}\label{anex:nonbalanced}
We shall consider a linear diattenuator illuminated by a non-balanced linear DFDP source, whose instantaneous Stokes vector is given in Eq.~(\ref{eq:nonbalancedlinear}). The output intensity shows the following DC and in-phase AC components:
\begin{gather}
I_{outL}^{0} = \rho I_0 + d{1-\gamma \over 1+\gamma}\cos(2\phi),\\
I_{outL}^{\Delta\omega X} = 2\rho I_0 {\sqrt{\gamma} \over 1+\gamma} d\sin(2\phi).
\end{gather}
It can be shown that the calculation of the OBC parameter according to Eq.~(\ref{eq:obc}) does not provide useful information for characterizing the sample dichroism, as the sample parameters $\rho$, $d$, and $\phi$ are strongly mixed.
If a non-balanced circular DFDP source is instead used, the DC, in-phase and quadrature components of the output intensity are:
\begin{gather}
I_{outC}^{0} = \rho I_0, \qquad I_{outC}^{\Delta\omega X} = I_{outL}^{\Delta\omega X},\\
I_{outC}^{\Delta\omega Y} = -2 \rho I_0 d {\sqrt{\gamma} \over 1+\gamma} \cos(2\phi).
\end{gather}
In this case, the OBC and the beatnote signal phase are respectively:
\begin{equation}
\mathrm{OBC}_C = 2d {\sqrt{\gamma} \over 1+\gamma}, \quad \text{and,} \quad \angle{I_{outC}^{\Delta\omega}} = 2\phi-\pi/2.
\end{equation}
It should be noted that the OBC is directly the diattenuation coefficient if $\gamma=1$. Therefore, a balanced source constitutes the most appropriate choice for our purposes. Regarding the phase of the beatnote signal, it enables the linear dichroism orientation to be readily obtained, as discussed in Section~\ref{sec:freespaceob}.

\section{}\label{anex:obellip}
We shall now consider a sample with elliptic dichroism, whose Mueller matrix $\mathbf{M_{ED}}$ has been derived in Appendix \ref{anex:muellerellip}. Using a linear DFDP source, the output is:
\begin{gather}
I_{outL}^{0} = \rho I_0, \\
I_{outL}^{\Delta\omega X} = \rho I_0 d \cos(2\epsilon) \sin(2\phi) , \\
I_{outL}^{\Delta\omega Y} = \rho I_0 d \sin(2\epsilon),
\end{gather}
which results in the following parameters:
\begin{gather}
\mathrm{OBC}_L=d \sqrt{\sin^2(2\epsilon) + \cos^2(2\epsilon)\sin^2(2\phi)}, \\
\angle{I_{outL}^{\Delta\omega}} = \arctan \left[ {\tan(2\epsilon) \over \sin(2\phi)} \right].
\end{gather}
Using a circular DFDP source, the results are:
\begin{gather}
I_{outC}^{0} = \rho I_0, \quad I_{outC}^{\Delta\omega X} = I_{outL}^{\Delta\omega X}, \\
I_{outC}^{\Delta\omega Y} = -\rho I_0 d \cos(2\epsilon) \cos(2\phi),
\end{gather}
which leads to:
\begin{gather}
\mathrm{OBC}_C=d \vert \cos(2\epsilon) \vert, \\
\angle{I_{outC}^{\Delta\omega}} = 2\phi-\pi/2.
\end{gather}
We can highlight two aspects of this configuration. The first one is
that the measured diattenuation coefficient $d$ diminishes as a
function of the dichroism ellipticity, completely vanishing in the
case of circular dichroism. The second one is that the phase of the
beatnote signal is the same regardless of the sample ellipticity, so it can
be extracted in the same way as in Eq.~(\ref{eq:ldangle}).

\section{}\label{anex:wgd}
Let $\theta$ be a random variable distributed according to a
Wrapped-Gaussian distribution (WGD) with probability
density function (pdf)~\cite{mardia99}
\begin{equation}
f_{WG}(\theta; \mu,\sigma)=\frac{1}{\sqrt{2\pi} \sigma}\sum_{k=-\infty}^{+\infty} e^{-\frac{\bigl(\theta-\mu+2k\,\pi\bigr)^2}{2\sigma^2}},
\end{equation}
where the parameters $\mu$ and $\sigma$ respectively identify with the mean and standard deviation of $\theta$.
Such a WGD verifies the following property:
\begin{equation}
\langle z^n\rangle=\int_\Gamma e^{in\theta} f_{WG}(\theta; \mu,\sigma)d\theta=e^{in \mu}e^{-\frac{n^2\,\sigma^2}{2}},
\end{equation}
where $z=e^{i\theta}$, and $\Gamma$ is an integration interval of length $2\pi$. 
As a result, the first moments of $z$ are
thus $\langle z \rangle=e^{i \mu}e^{-\frac{\sigma^2}{2}}$ and
$\langle z^2 \rangle=e^{2i \mu}e^{-2\sigma^2}$ and one has 
\begin{equation}
 \begin{gathered}
  \langle \sin \theta \rangle= e^{-\frac{\sigma^2}{2}}\sin \mu,\\
  \langle \cos \theta  \rangle= e^{-\frac{\sigma^2}{2}}\cos \mu.
  \end{gathered}
\end{equation}

\section{}\label{anex:cloude}
The Cloude coherency matrix $\mathbf{CM}$ of a Mueller matrix $\mathbf{M}$ can be
straightfully derived from the relations given in~\cite{cloude86}: $ \mathbf{CM}=(\sum_{j,k=1}^4  \mathbf{M}_{jk} \ \boldsymbol{\eta}_{jk})/4,$ with $\boldsymbol{\eta}_{jk}=\boldsymbol{\mathrm{T}}\bigl(\mathbf{\sigma}_j \otimes \mathbf{\sigma}_k^*\bigr)\boldsymbol{\mathrm{T}}^\dagger$, where the $\boldsymbol{\sigma}_{i,\,i\in[1,4]}$ stand for the standard Pauli
matrices, and where $\boldsymbol{\mathrm{T}}$ is given in Eq.~(\ref{eq:matrixt}). Using these relations, one obtains the Cloude
coherency matrix of $ \mathbf{M_{LD}^\Delta} $ (Eq.~(\ref{eq:mldrandom})), as 
\begin{equation}
\mathbf{CM_{LD}^\Delta}=\rho
\begin{bmatrix}
 \mathbf{CM_{LD,3\times3}^\Delta} & \vec{0} \\
\vec{0}^{\mathrm{T}} & 0
\end{bmatrix},
\end{equation}
where the upper $3\times 3$ matrix $\mathbf{CM_{LD,3\times3}^\Delta}$ reads
\begin{equation}
\rho \begin{bmatrix}1\text{+}\langle T \rangle & \bar{d}\, C_{2\bar{\phi}}\,e^{-2\sigma_\phi^2} & \bar{d} \,S_{2\bar{\phi}}\,e^{-2\sigma_\phi^2}\\
    \bar{d}\, C_{2\bar{\phi}}\,e^{-2\sigma_\phi^2}&\frac{1\text{-}\langle T \rangle}{2} \bigl[1\text{+}C_{4\bar{\phi}}\,e^{-8\sigma_\phi^2}\bigr]&\frac{1\text{-}\langle T \rangle}{2} S_{4\bar{\phi}}\,e^{-8\sigma_\phi^2} \\
    \bar{d}\, S_{2\bar{\phi}}\,e^{-2\sigma_\phi^2}&\frac{1\text{-}\langle T \rangle}{2} S_{4\bar{\phi}}\,e^{-8\sigma_\phi^2} &\frac{1\text{-}\langle T \rangle}{2} \bigl[1\text{-}C_{4\bar{\phi}}\,e^{-8\sigma_\phi^2}\bigr]\end{bmatrix}.
\end{equation}
%\end{widetext}
It can be clearly seen that this matrix is of rank three as soon as
$\sigma_\phi \neq 0$, whereas it is of rank one (independently of
$\sigma_\phi$) when $\langle T \rangle=1$ (isotropic case) thus leading to null Cloude entropy (${\cal S}=0$).

\section*{Acknowledgments}
This work has been funded by the French National Defense Agency (DGA)
and National Research Agency (ANR) project RADIO LIBRE
(ANR-13-ASTR-0001) and by R\'egion Bretagne.

\end{document}